\documentclass[12pt,preprint]{aastex}

\shorttitle{Modification To The Pressure Gradient}
\shortauthors{Huiquan Li & Jiancheng Wang}

\begin{document}


\title{A Modification To The Gaseous Pressure Gradients In Spherical
Celestial Bodies: Effects Of The Thermal Motion}

\author{Huiquan Li}

\affil{National Astronomical Observatories, Yunnan Observatory,
Chinese Academy of Sciences, P.O. Box 110, Kunming, Yunnan
Province 650011, P.R. China} \affil{Graduate School of the Chinese
Academy of Sciences. Email: lihuqu@yahoo.com.cn.}


\begin{abstract}
The atmospherical pressure gradients of spherical celestial bodies
are very important in astrophysics and other related areas. The
classical form of the gradient in a gravitation-dominated
celestial body: $\frac{dp}{dR}=-\frac{GM}{R^2}\rho$ has been
widely adopted for quit a long time. But, in this form, a factor
has been neglected: the centrifugal force that due to the thermal
motion of gaseous particles in a centripetal force field. The
pressure gradient is actually caused by the ``effective weight''
$f_g^{^{\prime }}$ of gaseous particles but not by their real
gravity $f_g$. For this reason, a modified form of the pressure
gradient is given in this paper. Though the changes are very small
in most celestial bodies, they may be not negligible in theories
of stars, accretion disks, and so on with a cumulative effect.
\end{abstract}



\section{Introduction}

It is well known that, in a spherical celestial body, the total
force density $F_R$ in the radial direction is the reason that
causes a gaseous pressure gradient along the radius $R$ (for
example to see \citet{lan80}):
\begin{equation}
\frac{dp}{dR}=-F_R.
\end{equation}
The force density $F_R$ is the summation of radial force
densities, which is possibly owing to different causes, e.g., the
gravity, the magnetic tension and the centrifugal force for the
celestial body rotation. The classical form of the atmosphere
pressure gradient in a spherical celestial body, such as the star
and the
planet, is often gravitation-dominated and expressed as: $F_R=\frac{GM}{R^2}%
\rho $, where $M$ is the total mass of the central object and
$\rho $ is the mass density of the gas. This formula can be
obtained with a simple mechanical means (though there are other
means that may be more complex): The pressure discrepancy of two
gaseous layers in the radial direction is caused by the gravity of
the gas between the two layers, which keeps a static balance of
the atmosphere. The extra pressure of the two layers is used to
support the intermediate gas.

However, we are not careful enough previously when using the
gaseous weight density in the spherical celestial bodies. Because
the gaseous particles are not static (contrarily they move with
very high thermal velocities), their effective weight
$f_g^{^{\prime }}$ that causes the pressure gradient is usually
smaller than the real weight $f_g$ (in section 2). So there is
actually a flatter pressure gradient in the spherical celestial
body than the previously expected classical form. In section 3, a
modified form of the gradient is given based on two ideal
experiments.

\section{The ``effective weight'' of moving particles in a spherical
celestial body}

We can define the ``effective weight'' in the case of a
``macroscopical'' object. When the object moves with a velocity
$\upsilon $ on the surface (a sphere) of a celestial body, it will
be balanced by three forces: the supporting force of the surface,
the gravity $f_g=\frac{GMm}{R^2}$ and the centrifugal force
$f_c=\frac{m\upsilon _{\perp }^2}R$, where, $M$, $m$ are the mass
of the object and the celestial body, and $\upsilon _{\perp }$ is
the velocity component that perpendicular to the radius $R$. For
the object, the centrifugal force $f_c$ must be equal to or
smaller than the gravity $f_g $ to keep it moving on the surface.
Otherwise, if $f_g<f_c$, the object will fly away from the
surface. In the case of $f_g\geq f_c$, we can get the effective
weight: $f_g^{^{\prime }}=f_g-f_c$, which is just equal to the
supporting force of the surface of the celestial body. It means
that, in a spherical celestial body, a moving object with velocity
component $\upsilon _{\perp }\neq 0$ always has an effective
weight, which is smaller than its real gravity. At this time, if
an external force that just $= f_g^{^{\prime }}$ affected on the
object in the locally radial and outward direction, then the
object can be balanced and will move in a circle of the celestial
body center. It also seems that a moving object in the centripetal
gravitation field has a smaller ``weight'' than in the case of
$\upsilon _{\perp } =  0$ or of a parallel force field.

As easily extended, the same thing will happen not only for
``macroscopical'' objects but also for hot particles in stars and
accretion disks, though the moving directions of these particles
are random. Previously, when we considered the atmospheric
pressure gradient, we set the gaseous particles as a system (the
system is statistically static) but not took it account of the
particle motion in the system.

In fact, based on the balance condition, the atmospheric pressure
gradient of the celestial body is caused by the effective weight
$f_g^{^{\prime }}$ but not by the real weight $f_g$ of the gaseous
particles. For many celestial bodies with very high temperatures
(can reach about $10^6$ K), the thermal speeds of gaseous
particles are usually very fast. From the above analysis, we can
infer that these particles should have smaller effective weight.
And a gaseous system composed of these particles should also have
a smaller total effective weight than its real total gravity. So
the hot gas between two layers needs a smaller supporting force
from the down layer. That is to say, the gaseous pressure gradient
should be flatter than the classical form. What is more, this
result is away right no matter what the atmosphere is composed of
(most possibly can be plasma), because the pressure gradient is
determined by the effective gravity density but has nothing to do
with the interactions between particles (atoms, molecules or
ions).

\section{The modified pressure gradient in a celestial body}

Firstly, we shall analysis the mechanism of every individual
particle (assumed to be the same sort of particles with a mass
$m$) in a spherical and gas-filled celestial body, only
considering the radial gravity $f_g$ and ignoring other possible
centripetal forces. In supposition, there are two equivalent
cases: (i) If the gravity $f_g=0$ in the spherical space, there is
no difference of gaseous properties from in a quadrate container.
The pressure gradient of the gas is zero everywhere. (ii)
Otherwise, if there is a centripetal force $f_g\neq 0$, which is
different for different particles and happens to be equal to the
centrifugal force $f_c$ that due to the thermal motion of the
corresponding particle, the pressure gradient of the gas is also
zero like in the case (i). Although it is impossible to keep the
condition $f_g=f_c$ always true for different particles (their
thermal velocities are different) when the mass of the central
object is given, these two equivalent cases provide some clues on
which the centrifugal force of the thermal motion is able to
affect the atmosphere pressure gradient when $f_g\neq 0$.

A general relation between $f_g$ and $f_c$ can only be $f_g\geq
f_c$, because, if the centrifugal force is larger than the
gravity, the particle bounded in a gravitational potential will
escape from the region where they
are located. In the following, we shall discuss the pressure gradient when $%
f_g\geq f_c$ as existing in general celestial bodies (see examples
at the last paragraph).

For an arbitrary particle $i$ in an infinitely small shell (a unit
volume) at the radius $R$, where there are $n$ particles (also is
the number density of the gas), the thermal velocity is in an
arbitrary direction and the centrifugal force $f_c(i)$ is
connected with the thermal velocity
component $\upsilon _{\perp }(i)$ that perpendicular to the radius: $f_c(i)=%
\frac{mv_{\perp }^2(i)}R$. The component $\upsilon _R(i)$ in the
direction along the radius brings no centrifugal force. If
$f_c(i)\leq f_g(i)$, we can divide the gravity $f_g(i)$ into two
parts. One supplies the restriction against the departure trend
from its ``circular''\ trajectory in the commoving frame of
$\upsilon _R(i)$, which is equal to $f_c(i)$ and so make the
physical properties in the small spherical shell equivalent as in
a quadrate container. In this container, the remnant (effective)
force acting on the particle $i$ in the radial direction is
$f_R(i)=f_g(i)-f_c(i)$. So the summated effective force of $n$
particles in the unit quadrate volume along the vector
$\overrightarrow{R}$ is:
\begin{equation}
F_R=\sum_{i=1}^nf_R(i)=\frac{GM}{R^2}\rho
-\sum_{i=1}^n\frac{m\upsilon _{\perp }^2(i)}R=\frac{GM}{R^2}\rho
-\frac{nm\left\langle \upsilon _{\perp }^2\right\rangle }R,
\end{equation}
which is also the total force density on the particles in the
spherical shell equivalently.

In the system, the ``effective weight'' $f_R(i)$ doesn't change
with the time and collisions. In fact, the locally transverse
component of the particle velocity is kept the same absolute value
(when $f_g\geq f_c$) no matter which place it will move to if not
considering the collision with other particles. Because the
gravity does not accelerate or decelerate the absolute value of
the locally transverse velocity, but only changes its direction
and the kinetic energy in the radial direction. Even when
collisions happen, the particles only communicate their velocities
in the case of the same mass. At different time, the state of the
particle system is invariable as a whole.

The velocity $\upsilon $ can be decomposed in three directions $x$, $y$ and $%
z$. If firstly the effective force density, which only gives a
pressure gradient, is not considered, a statistical relation
connects the averaged square values of these decomposed
velocities: $\left\langle \upsilon _x^2\right\rangle =\langle
\upsilon _y^2\rangle =\left\langle \upsilon _z^2\right\rangle
=\frac 13\left\langle \upsilon ^2\right\rangle $. The square
velocity perpendicular to the radius that averaged on all
particles is: $\left\langle \upsilon _{\perp }^2\right\rangle
=\left\langle \upsilon ^2\right\rangle -\left\langle \upsilon
_R^2\right\rangle =\frac 23\left\langle \upsilon ^2\right\rangle
$. In thermodynamics, the local
pressure describes the averaged kinetic energy of particles in this way: $%
p=\frac 13nm\left\langle \upsilon ^2\right\rangle $. So the total
force density in the radial direction is: $F_R=\frac{GM}{R^2}\rho
-\frac{2p}R$,
where the mass density $\rho $ is confined by the equation of a perfect gas $%
p=\frac \rho mkT$. The pressure gradient by taking account of the
centrifugal force for the thermal motion of particles should be
modified to this form:
\begin{equation}
\frac{dp}{dR}=-(\frac{GMm}{kTR^2}-\frac 2R)p.
\end{equation}
For the centrifugal force, it seems to be an odd point at the object center $%
R=0$. But, in fact, this odd point doesn't really exist. In any
static celestial bodies, the centrifugal force must satisfy the
necessary condition that the gaseous particles are restricted:
$f_g(R)\geq f_c(R)$. Otherwise, the particles will move to outer
regions of the object and are in an imbalance state. For example,
in a purely gaseous system, the self-gravity in the central region
will be not large enough to bound the particles. The particles
will flee away from the gravitational center. In this system, the
gravity tend to just provide the restriction on the gaseous
particles and brings no pressure gradient. So the matter at the
center of a general celestial body is usually condensed or
compact.

In the right side of the above formula, the relative magnitude of
the second
term to the first term can be represented by the parameter $\chi =\frac{2kTR%
}{GMm}$ ($\chi $ always $\leq 1$). The importance of the second
term depends on the atmospheric temperature of the celestial body.
On the ground of the earth, $\chi $ is about $2.7\times 10^{-3}$;
Even in an altitude $\sim 10^3$ Km from the ground, the
temperature is $\sim 10^3$ K and $\chi \sim $ $0.01$
\citep{hed83}. For the sun, the parameter $\chi $ is some larger
in some regions such as in the corona $(\chi \sim $ $0.087$) and
below the convection zone. But the thermal centrifugal force is
possibly important in some fields such as the accretion disk, the
star formation and evolution scenario.

\acknowledgments

This work was supported in part by National Natural Science
Foundation of China (NSFC) 0012AA and by the Special Funds for
Major State Basic Science Research Projects of China 1303 AA.

\end{document}